\documentclass{nature}
\usepackage{textcomp}
\usepackage{siunitx}
\usepackage[pdftex]{graphicx}

\usepackage[colorlinks]{hyperref}
\usepackage{graphicx, amsmath, color}
\usepackage[normalem]{ulem}
\makeatletter
\let\saved@includegraphics\includegraphics
\AtBeginDocument{\let\includegraphics\saved@includegraphics}
\renewenvironment*{figure}{\@float{figure}}{\end@float}
\usepackage{amssymb}
\makeatother
\usepackage[colorlinks]{hyperref}

\usepackage{dcolumn}
\usepackage{bm}
\usepackage{gensymb}

\title{Attosecond synchronization of extreme ultraviolet high harmonics from crystals}

\author{Giulio Vampa$^1$, Jian Lu$^1$, Yong Sing You$^1$, Denitsa R. Baykusheva$^1$, Mengxi Wu$^2$, Hanzhe Liu$^1$, Kenneth J. Schafer$^2$,  Mette B. Gaarde$^2$, David A. Reis$^1$ \& Shambhu Ghimire$^1$}

\begin{document}

\maketitle

\begin{affiliations}
\item Stanford PULSE Institute, SLAC National Accelerator Laboratory, Menlo Park, California 94025, USA.
\item{Department of Physics and Astronomy, Louisiana State University, Baton Rouge, Louisiana 70803, USA}
\end{affiliations}

\begin{abstract}

    The interaction of strong near-infrared (NIR) laser pulses with wide-bandgap dielectrics produces high harmonics in the extreme ultraviolet (XUV) wavelength range. These observations have opened up the possibility of attosecond metrology in solids, which would benefit from a precise measurement of the emission times of individual harmonics with respect to the NIR laser field. Here we show that, when high-harmonics are detected from the input surface of a magnesium oxide crystal, a bichromatic probing of the XUV emission shows a clear synchronization largely consistent with a semiclassical model of electron-hole recollisions in bulk solids. On the other hand, the bichromatic spectrogram of harmonics originating from the exit surface of the 200 $\mu$m-thick crystal is strongly modified, indicating the influence of laser field distortions during propagation. Our tracking of sub-cycle electron and hole re-collisions at XUV energies is relevant to the development of solid-state sources of attosecond pulses.
    
\end{abstract}

\section{Introduction}
Generation of extreme ultraviolet (XUV) high harmonics from gaseous media has been the foundation of attosecond science \cite{corkum2007attosecond, krausz2009attosecond}, which includes attosecond pulse generation \cite{paul2001observation}, imaging molecular orbitals \cite{itatani2004tomographic} and attosecond tunneling interferometry \cite{shafir2012resolving, pedatzur2015attosecond}. At the heart of atomic high-order harmonic generation (HHG) lies a three-step recollision process  \cite{corkum1993plasma} that consists of tunnel ionization, free-electron acceleration, and recollision to the parent ion. Based on their kinetic energies, electrons recollide with the parent ions at slightly different times in the subcycle scale\cite{dudovich2006measuring}, causing an intrinsic delay between harmonics. This delay, termed ``atto-chirp", is deleterious for crafting transform-limited isolated attosecond pulses or attosecond pulse trains. The precise timing between the electron trajectories is the cornerstone of high-harmonic spectroscopy\cite{shafir2012resolving}. Following the recent observation of high-harmonic emission from bulk crystals \cite{ghimire2011observation, schubert2014sub, solidargon, hohenleutner2015real, vampanature, luu2015extreme, garg2016multi, liu2017high, you2017anisotropic}, attosecond metrology is being extended to solids, with methods developed to reconstruct electronic band structures in reciprocal space \cite{vampanature,vampa2015all,luu2015extreme}, to probe the periodic potential in real space \cite{you2017anisotropic,liu2017high}, and to boost the emission efficiency in nano-structures \cite{han2016high,sivis2017tailored,vampa2017plasmon,liu2018enhanced}, as well as towards stable attosecond pulses \cite{garg2018ultimate}. Just like in gas-phase HHG, many of these applications benefit from the understanding of the temporal connection between harmonics and the driving NIR laser field at the sub-cycle level. 

In solids, there are two major HHG channels and they are expected to have distinct temporal profiles\cite{wu2015high}.  XUV harmonics from thin SiO$_2$  subjected to NIR laser fields were found chirp-free \cite{luu2015extreme,PhysRevA.98.041802}, which is consistent with Bloch-like nonlinear oscillations of electrons in the conduction bands. The other competing channel  is inter-band polarization or recollision based harmonics\cite{vampanature,vampa2014theoretical,vampa2015semiclassical,wu2015high}, similar to HHG from gases. As shown in ultraviolet harmonics from ZnO, this process preserves the temporal mapping characteristic of gas-phase HHG\cite{vampanature}. In MgO crystals, the modulation of the high-harmonic spectrum with the carrier-envelope phase of the few-cycle laser pulse \cite{you2017}, as well as the laser-intensity-induced shift in the emission phase of individual harmonics\cite{lu2019interferometry} also point towards the inter-band or recollision-based emission, however the atto-chirp has not been measured. 

Here we apply a bichromatic probing scheme\cite{vampanature,dudovich2006measuring,pedatzur2015attosecond,shafir2012resolving} to XUV high-harmonics emitted from the input and exit surfaces of a magnesium oxide crystal. Harmonics emitted from the input surface show clear spectral signatures consistent with recolliding electron-hole pair trajectories, and quantify the ``atto-chirp''. This is our first result. Spectrograms of harmonics emitted from the exit surface of a 200 micrometer thick MgO, however, are strongly distorted. This is our second result. Together with a measured broadening and blue-shift of the transmitted NIR spectrum, we conclude that the bichromatic probing scheme encodes temporal nonlinearities experienced by the NIR pump during propagation through the crystal. Spatial distortions of the XUV high-harmonic beam from the same crystal has been recently reported, too\cite{vampa2018observation}. Lastly, we develop a quantum-mechanical model of high-harmonic generation in solids that adds tunnelling of the electron-hole pair across the minimum bangap of MgO. Implications of our work include  the possibility of developing attosecond metrologies based on recolliding electrons at XUV photon energies in solids and the generation of attosecond pulses with crystals. 

\section{Experiment}
In the experiment, we measure high harmonics from MgO crystals with 200 $\mu$m thickness subjected to a NIR field centered at 1320 nm and its weak second harmonic. Both the fundamental and second harmonic are polarized along the [100] direction of the crystallographic axis. We record the XUV high-harmonic spectra as a function of the attosecond delay between the two colors, which is controlled with a pair of glass wedges (see Supplementary Information for detail).

\begin{figure}
    \centering
    \includegraphics[scale = 1.6]{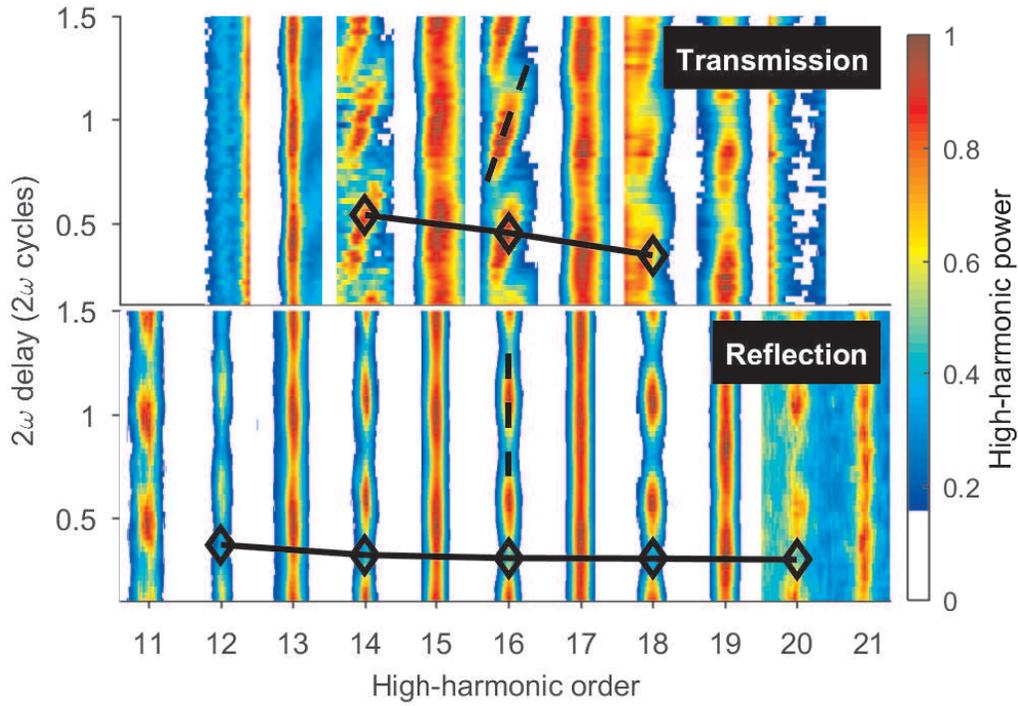}
    \caption{The modulation of the even harmonics measured in the transmission geometry (top panel) shows an intra-harmonic chirp (dashed black line), resulting in inconsistent determination of the modulation phase. Unequivocal determination of the modulation phase is possible, instead, in the reflection geometry - here set at a 45\degree angle of incidence (bottom panel). The local minima for the even-harmonic modulation are extracted and indicated by the diamond symbols. 
    Each harmonic is independently normalized to 1.}
    \label{fig:spectrograms}
\end{figure}

We begin our investigation by showing the relevance of temporal distortions of the infrared field upon propagation. This is performed by comparing two-color driven harmonic spectra in the transmission mode (Fig. \ref{fig:spectrograms}, top panel) with those in the reflection mode (Fig. \ref{fig:spectrograms}, bottom panel), but otherwise under similar conditions (See Supplementary Information for details). The excitation intensity inside the sample in both cases (considering Fresnel loss) is estimated to be $\sim$10 TW/cm$^2$. Because XUV harmonics are effectively emitted within a thickness that is on the order of one absorption length ($\sim$ 10 nm), from the entrance side in the reflection geometry and from the exit side in the transmission geometry, the transmitted harmonics are expected to encode the distortions accumulated by the pump pulse during propagation from the entrance to the exit side of the sample. Indeed, transmitted harmonics exhibit broader peaks, which we attribute to the broader (by about 30 percent) and blue-shifted spectrum of the fundamental pulse upon propagation (See Supplementary Information). Moreover, each individual transmitted harmonic exhibits a linear frequency shift with the second harmonic delay (shown by the dashed line), which is consistent with different NIR center frequencies across the pulse. Spectral broadening and frequency shifts are a result of nonlinear propagation effects, possibly self-phase modulation, since no frequency shift on individual harmonics is measured in reflection mode (dashed line). Linear dispersion of MgO is ruled out as a possible contribution by adding a similarly thick MgO crystal in the beam path before the focusing lens, for the reflection geometry.

In addition to the individual frequency shifts, there is a delay across neighbouring harmonics. As shown in a previous work \cite{dudovich2006measuring}, this modulation provides a measurement of the sub-cycle emission time of the harmonics, the so-called ``atto-chirp". This quantity is derived by matching the observed modulation with that predicted by a model of recolliding electron-hole pairs in MgO, as described in the next paragraph and in the Supplementary Information.
In the reflection geometry, we measure a lower limit for the attochirp decreasing from $148 \pm 44$ as/eV at $\sim 11$ eV to $11 \pm 44$ as/eV at $\sim 15$ eV, when a semiclassical model of recolliding electron-hole pairs is considered. The atto-chirp deduced with a quantum model (described below) is compatible with the semiclassical one, but has a larger uncertainty (Supplementary Figure 4). The phase of the inter-harmonic modulation is stronger in the transmission geometry, but the marked intra-harmonic shift renders determination of this phase inconsistent. 

\begin{figure}
    \centering
    \includegraphics[scale = 1.6]{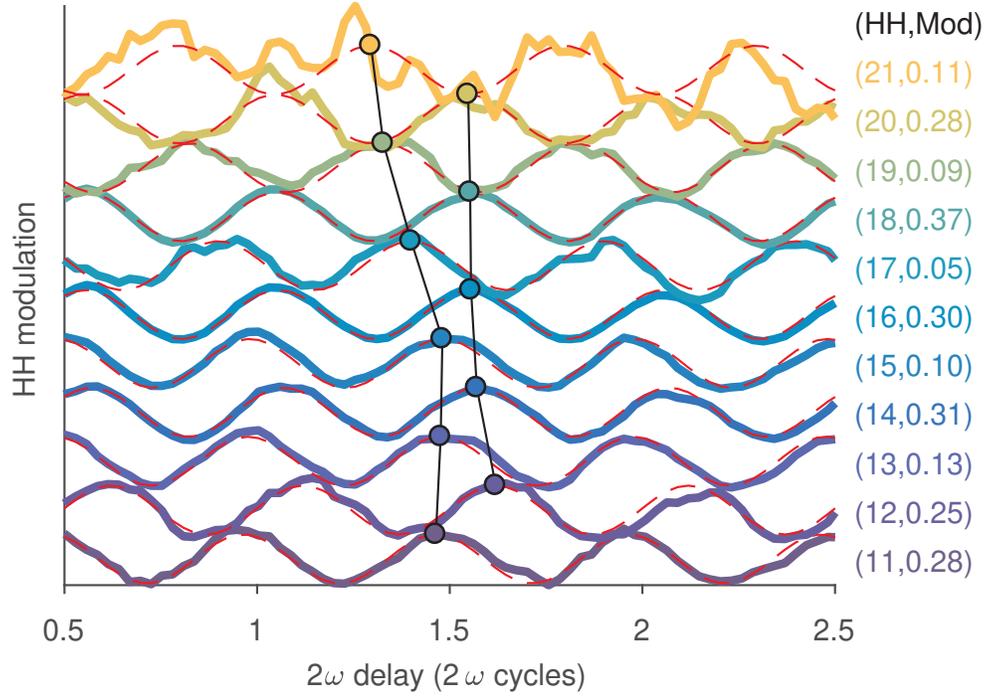}
    \caption{Modulation of high-harmonic power for harmonics $11^{th}$ to $21^{st}$ (color coded) versus sub-cycle delay between the fundamental and second harmonic fields (in cycles of the second harmonic) measured in the reflection geometry. The order of the harmonic and the modulation amplitude (normalized to 1) are reported in parentheses next to the curves on the right hand side. The delay that yields the highest power is marked by colored circles for every harmonic. The dashed red lines are fits to the experimental modulation with function $\cos(\phi+\phi_{opt})$, where $\phi$ is delay. The only fit parameter is the offset phase, $\phi_{opt}$. The second harmonic power is set to $\sim$0.5\% of the fundamental. 
    } 
    \label{fig:modulations}
\end{figure}

Next, we analyze the observed modulation of the even-order harmonics as a function of the two-color delay in detail, for the reflection geometry. When a second harmonic field with $\sim$0.5\% of the power of the pump is added to the fundamental driver and properly phased, the  asymmetric field breaks the inversion symmetry of the high-harmonic dipole, resulting in electrons and holes being accelerated farther apart or closer together at subsequent laser half-cycles of the driver. The uneven path length is described by the addition or subtraction of a phase $\sigma(n\omega,\phi)$ to the oscillating high-harmonic dipole \cite{dudovich2006measuring,vampanature}, dependent on the harmonic order $n$ and two-color delay $\phi$ (see Supplementary Information). As the delay $\phi$ is varied, the high-harmonic power modulates with an order-dependent phase. This is shown in Fig. \ref{fig:modulations} for the reflection case.

\begin{figure}
    \centering
    \includegraphics[trim={2.5in, 1.5in, 1.5in, 1.5in}]{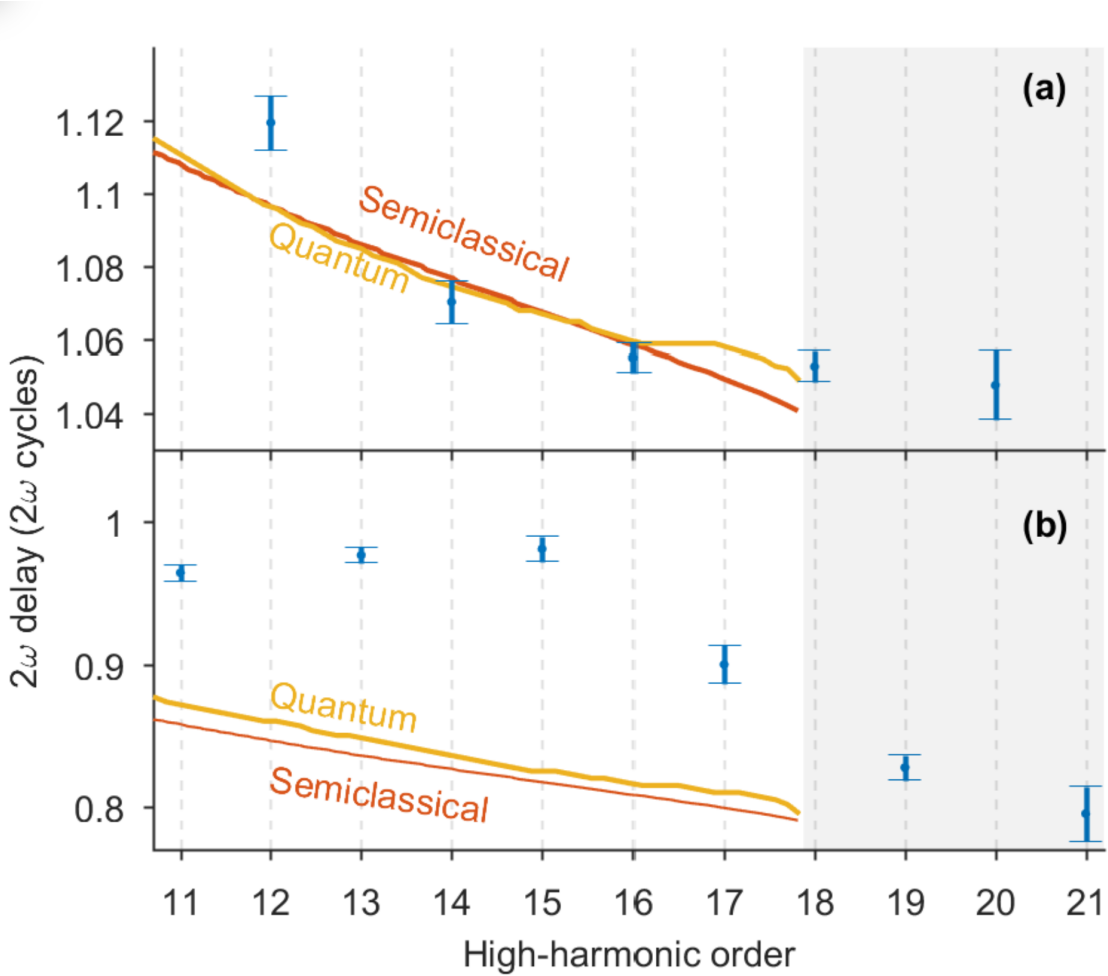}
    \caption{(a) Both the semi-classical and quantum correction models of recolliding electron-hole pairs predict a harmonic order dependent phase for the modulation of the even-order harmonics. They closely match the experimental data (blue markers), up to an offset (see Supplementary Information for details). This is the so called ``atto-chirp". The odd harmonics (panel b) also show order dependence but they deviate significantly from the model. The predictions from the model are restricted within the maximum band-gap at the zone edge. The 18$^{th}$ and 20$^{th}$ harmonics lie above the maximum band-gap, indicated by the grey area.}
    \label{fig:phasecomparison}
\end{figure}

Experimental parameters are reported in the Supplementary Information. The delay that yields the highest harmonic power ($\phi_{opt}$) is extracted from a cosine fit of the normalized modulations with a fixed frequency for all harmonics. It is plotted in Fig. \ref{fig:modulations} (colored circles) for the even and the odd harmonics separately, and in Fig. \ref{fig:phasecomparison}, where it is compared with the theoretical predictions. Overall, the delay for the even harmonics agrees reasonably well with the simple semiclassical three-step model introduced above (orange line) \cite{vampa2014theoretical,vampa2015semiclassical}, up to an unmeasured offset phase. In essence, the agreement suggests that XUV harmonics from MgO are a result of  recollisions between electrons and their associated holes that are driven by the strong laser fields in the lowest conduction and one of the highest valence bands, respectively. Bloch-like emission, instead, predicts a modulation phase which is either in-phase or out-of-phase with the second harmonic delay, but without atto-chirp (see derivation in the Supplementary Information). We note that the semiclassical model  predicts the emission phase of harmonics only up to 18$^{th}$ order because of the limit set by the maximum band-gap at the zone edge. The emission of harmonics beyond this order would require considerations of tunneling to a higher-lying conduction band \cite{you2017,you2019} although experimental data does not show any apparent abrupt changes in the emission phase around this energy range. 

We note that the semiclassical model neglects quantum aspects  similar to those studied both theoretically \cite{ivanov2014multielectron} and experimentally \cite{pedatzur2015attosecond} in the gas phase. Here, we extend those calculations to the solid-state platform for the first time (see derivation in the Supplementary Information), see yellow line Fig. \ref{fig:phasecomparison}a.  However, the difference is not statistically significant and is within the uncertainty in the extracted phase for otherwise fixed model parameters.  The theoretical framework that we have developed for solids predicts an imaginary component of the birth time of about 460 as for harmonics in the photon energy range from 12 eV to 18 eV (Supplementary Figure 1), corresponding to $\sim0.1$ cycles of the NIR field.  This value is similar to that measured in He atoms driven at 4.4 $\times 10^{14}$  W/cm$^2$ \cite{pedatzur2015attosecond}. However, we find a large offset of the modulation phase for the odd harmonics, compared to both the semiclassical model and that with this extended model (figure \ref{fig:phasecomparison}b). The large deviation might arise from physics beyond our model or can be a signature of a second-harmonic strength larger than desirable. Supplementary Figure 5 shows, in fact, that for increasing second-harmonic strength the relative modulation phase for adjacent even and odd harmonics progressively decreases. A slight decrease in the relative phase between the even harmonics is also observed as the second harmonic strength increases to 0.28\% of the fundamental. Therefore, the reported atto-chirps shall be considered lower limits.

\section{Conclusions}
In conclusion, we measured the attosecond synchronization of XUV harmonics from 200 micrometer thick MgO crystals subjected to intense NIR laser fields. The results obtained in the reflection geometry closely represent the intrinsic delay of high-harmonics predicted by generalized re-collision model, whereas two-color spectrograms of harmonics measured in the transmission geometry show strong temporal distortions - a cautionary tale for performing \textit{in-situ} high-harmonic spectroscopy in this geometry. Our evidence suggests that they are strongly influenced by propagation effects. In the reflection geometry, using semiclassical trajectories we extract a minimum atto-chirp of $11 \pm 44$ as/eV (lower limit) about the 16$^{th}$ harmonic at $\sim$15 eV. 
With proper dispersion compensation, such as with ultra-thin metal filters, XUV harmonics from MgO could support attosecond pulse trains. Typical gas-phase harmonic sources operate at  about two orders of magnitude higher laser intensities. Because of the modest peak intensity requirements, solid-state HHG based sources should be feasible with modern high repetition rate laser systems\cite{chini} such that the total XUV flux can be increased significantly.




\begin{addendum}
 \item[Acknowledgements] 
 {\flushleft At Stanford/SLAC, this work was primarily supported by the US Department of Energy, Office of Science, Basic Energy Sciences, Chemical Sciences, Geosciences, and Biosciences Division through the Early Career Research Program. G.V. acknowledges the support from W. M. Keck Foundation. D.B. acknowledges the support from the Swiss National Science Foundation (SNSF) through project number P2EZP2-184255. The work at LSU was supported by the National Science Foundation (NSF) (PHY-1713671). \\
 The authors declare that they have no competing financial interests.}
 \item[Author Contributions]
 S.G. conceived the experiments.  J. L. conducted experiments, G.V performed calculations. G.V. and J.L contributed  equally to this work. Y.S.Y conducted initial measurements. M.W. and D.B. also performed independent calculations. H.L. helped in initial measurements. All authors contributed to the interpretation of the results and writing of the manuscript. 
\end{addendum}

\flushleft{ 
\large{\textbf{References}}
\bibliographystyle{naturemag}
{\footnotesize
\bibliography{biblio}}
}

\end{document}



\title{Attosecond synchronization of extreme ultraviolet high harmonics from crystals\\
Supplementary Information}

\author{Giulio Vampa}
\email{gvampa@stanford.edu}
\author{Jian Lu}%
\author{Yong Sing You}%
\author{Denitsa R. Baykusheva}%
\author{Mengxi Wu}%
\author{Hanzhe Liu}%
\author{Kenneth J. Schafer}%
\author{David A. Reis}%
\author{Mette B. Gaarde}%
\author{Shambhu Ghimire}
\email{shambhu@slac.stanford.edu}

\affiliation{Stanford PULSE Institute, SLAC National Accelerator Laboratory, Menlo Park, California 94025, USA}
\affiliation{Department of Physics and Astronomy, Louisiana State University, Baton Rouge, Louisiana 70803, USA}


\date{\today}

\maketitle
\section{Experimental details}
\begin{figure}[h]
    \centering
    \includegraphics{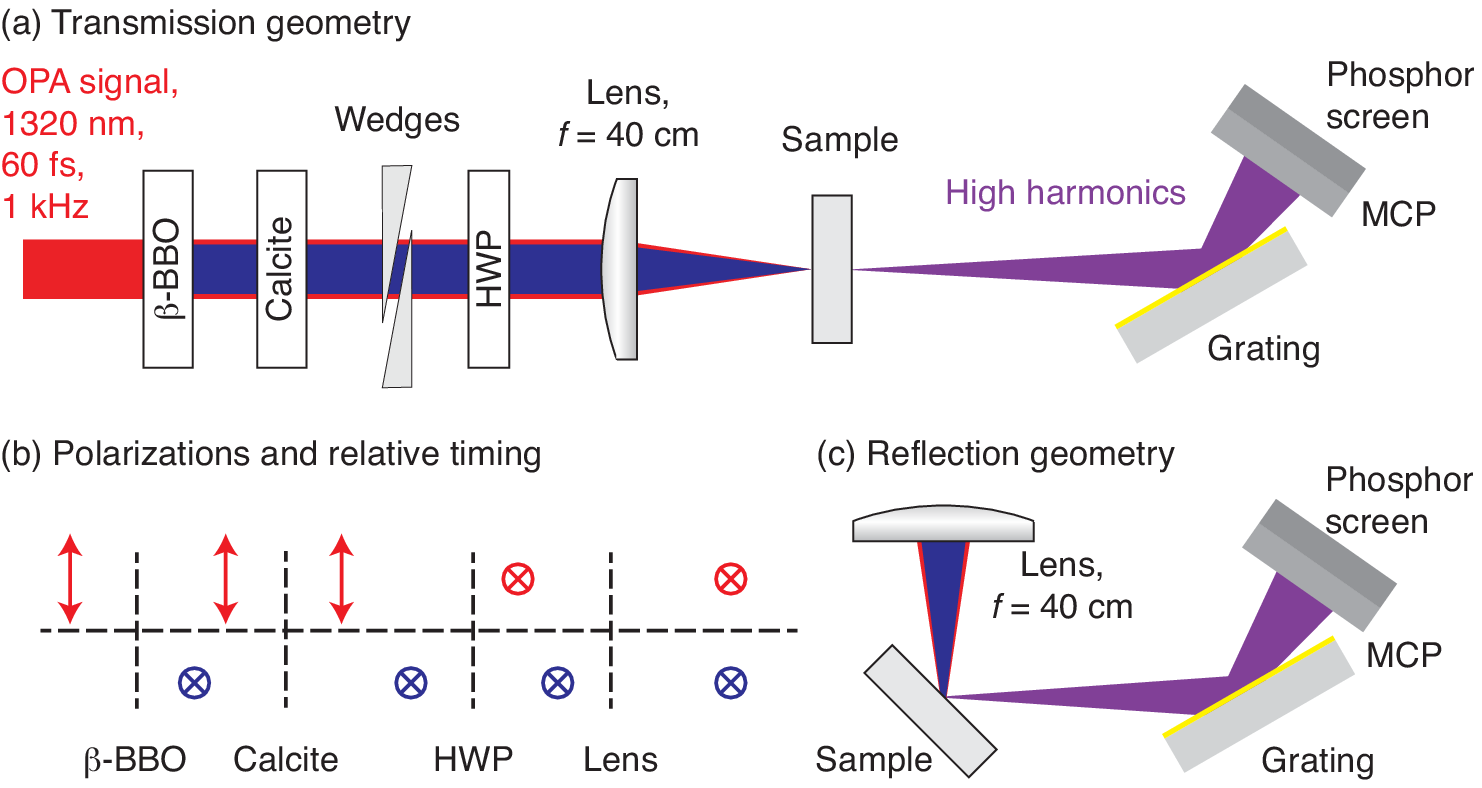}
    \caption{(a) Experimental setup in the transmission geometry. OPA: optical parametric amplifier; $\beta$-BBO: beta barium borate; HWP: half wave-plate; MCP: micro-channel plate. One wedge is translated using a motorized stage. (b) The polarization in the laboratory frame and the relative timing for the fundamental and second-harmonic pulses after transmission through each optics. (c) Experimental setup in the reflection geometry. The sample is rotated by 45$\degree$ with respect to the geometry in (a). The fundamental pulses at 1320 nm are indicated in red and the second-harmonic pulses are indicated in blue.} 
    \label{fig:setup}
\end{figure}

The experimental setup in the transmission geometry is shown in detail in Fig. \ref{fig:setup}(a). The NIR pulses at 1320 nm with $\sim$60 fs duration are generated from an optical parametric amplifier pumped by an amplified Ti:Sapphire laser system operating at a 1 kHz repetition rate. We superimpose a weak second-harmonic field to the intense driver using an interferometer in an inline geometry \cite{dudovich2006measuring}, which ensures sufficient phase stability. The second-harmonic field is generated with orthogonal polarization with respect to the NIR field using a 100 $\mu$m-thick beta barium borate crystal via type-I phase matching. A half wave-plate rotates the NIR polarization to be vertical in the laboratory frame, which is parallel to that of the second-harmonic field. The half wave-plate is zero-order for 1320 nm light and does not alter the polarization for the second-harmonic light at 660 nm. A lens with a focal length of 40 cm focuses the laser beams onto the sample. A 1 mm-thick birefringent calcite plate placed before the half wave-plate compensates the delay between the fundamental and second-harmonic pulses introduced by the transmissive optics in the beam path, such that the two-color fields are temporally overlapped at the sample. The phase delay between the two-color fields is controlled with attosecond precision by varying the insertion of a pair of glass wedges in the beam. The polarization for the fundamental and second-harmonic pulses and the relative timing between them after transmission through each optics are shown in Fig. \ref{fig:setup}(b). The transmission measurements are conducted at normal incidence while the reflection measurements are conducted at a 45$\degree$ angle of incidence and the harmonics propagating along the specular reflection direction are collected. The XUV high harmonics emerging from the sample are collected by an XUV spectrometer consisting of a grating (600 grooves per mm), a micro-channel plate, and a phosphor screen. The spectrographs displayed on the phosphor screen are recorded by an imaging lens and a charge-coupled device. The sample and the XUV spectrometer are held in vacuum to avoid air absorption of the XUV harmonics.

\begin{figure}[h]
    \centering
    \includegraphics{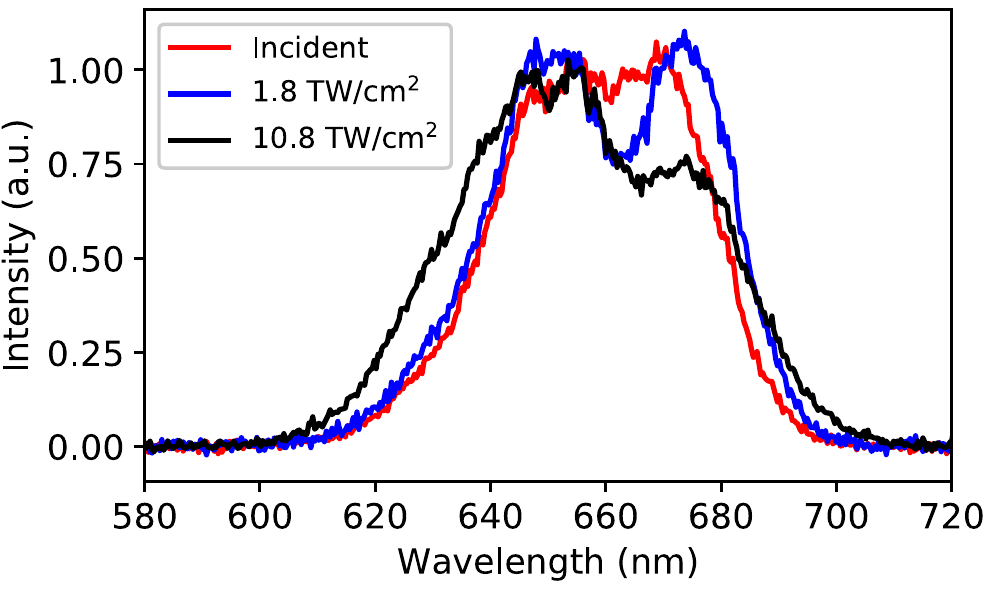}
    \caption{The second-harmonic spectrum of the pump pulse at 1320 nm (red) and second-harmonic spectra of the pump pulse after propagation in a 200 $\mu$m thick MgO sample at two different intensities (blue and black). The intensities inside the samples are indicated in the legend. The pump pulse polarization is parallel to the [100] axis. The pump spectrum undergoes significant spectral broadening after propagating through the sample, indicating the strong propagation effects in experiments in transmission geometry.} 
    \label{fig:propagation}
\end{figure}

To show the relevance of the propagation effects associated with the pump pulses, we measured the second-harmonic spectra of the fundamental pulses before and after propagation through the 200 $\mu$m MgO sample with polarization parallel to the crystal cubic axis. The second-harmonic spectra are shown in Fig. \ref{fig:propagation}. Significant broadening and spectral modulation of the second-harmonic spectrum after the fundamental pulse transmitting through the sample are evident. This suggests that the propagation effects of the strong pump pulses are non-negligible in our high-harmonic generation experiments in transmission geometry.

\section{Analysis of the optimum phase}

The lack of the dynamical inversion symmetry perturbs the high-harmonic dipole as follows:
\begin{align}
\label{eq:perturbeddipole}
d &=  {d_0 e^{-in\omega t -i\sigma}-d_0e^{-in\omega (t+\pi/\omega)+i\sigma}} = \\
&= d_0 \begin{cases} \sin{\sigma},&\mbox{for }n = 2m\\
\cos{\sigma}, & \mbox{for }n=2m+1 \end{cases}
\end{align}
where $d_0$ is the unperturbed high-harmonic dipole (inversion symmetric), $n$ is the harmonic order, $\omega$ is the frequency of the driver, and
\begin{equation}
    \label{eq:sigma}
    \sigma(t_r,\phi) = \int_{t_i(t_r)}^{t_r} \mathrm{\mathbf{v}}(\tau,t_i)\mathbf{A}_2(\tau,\phi)d\tau.
\end{equation}
Here, $\rm{\mathbf{v}(\mathbf{k})} = \rm{\mathbf{v}_c(\mathbf{k})}-\rm{\mathbf{v}_v(\mathbf{k})} = \mathbf{\nabla_{k}}\varepsilon_g(\mathbf{k})$
 is the semiclassical velocity difference between electrons and holes in the valence and conduction bands respectively, $\varepsilon(\mathbf{k})$ is the energy difference between the two bands, and $\mathbf{A}_2(\tau,\phi) = \mathbf{A}_2\cos{(2\omega\tau + \phi)}$ is the vector potential of the second harmonic. The additional phase is accumulated between the unperturbed times of creation of the electron-hole pair ($t_i$) and of recollision of the electron with the hole ($t_r$). As the delay $\phi$ between the driver and the second harmonic is varied, the high-harmonic power modulates according to Eq. (\ref{eq:sigma}) and the Fourier transform of Eq. (\ref{eq:perturbeddipole}), with an order-dependent phase.  The classical trajectories are calculated following the procedure described in Ref. \cite{vampa2014theoretical,vampa2015semiclassical} but using the band structure of MgO  \cite{you2017}. Uncertainty in the calculated bandstructure such as typical under estimation of the band-gap by DFT is discussed elsewhere \cite{you2017}. Quantum trajectories, instead, are calculated according to the procedure reported in the next section. Once the trajectories are known, Eq. (\ref{eq:sigma}) is evaluated numerically in either the real (classical) or complex (quantum) plane. Next, the harmonic modulation is calculated with Eq. (\ref{eq:perturbeddipole}). Finally, the phase of the modulation is extracted for each harmonic and plotted in Fig. (3) of the main text. Error bars in experimental results represent 95 percent confidence bounds obtained from the QR-decomposition procedure adopted in the nonlinear least squares regression. Because in experiment we do not resolve the absolute phase offset, we displace the theoretical curves by an amount that minimizes the value of reduced $\chi^2$. Here $\chi^2 = \sum (theory - experiment)^2/\sigma^2$, where $\sigma$ corresponds to the 68 percent confidence interval on the phase  modulation of individual harmonics. The value of reduced $\chi^2$ for semi-classical and quantum correction models are 23.36 and 23.96 respectively. We note that all calculation results are performed in one dimension assuming that the linearly polarized laser field is aligned along [100] direction. As shown recently through polarization measurements of XUV harmonics from MgO, slight deviations  from  high symmetry directions of the crystal require considering anisotropic two dimensional effects \cite{you2019}.   

\section{Solving for the quantum orbits}
The quantum mechanical model is developed to calculate the dominant quantum orbits of the high-harmonic dipole path integral, Eq. (4b) of Ref. \cite{vampa2014theoretical}. The orbits are characterized by complex times of creation of electron-hole pairs and their annihilation (the imaginary component encoding quantum-mechanical tunnelling), as well as complex momenta \cite{dahlstrom2011quantum,ivanov2014multielectron}. The derivation is reported below. 
The quantum orbits are derived by finding the conditions of least action with saddle point integration. The saddle point equations are \cite{vampa2014theoretical}:
\begin{subequations}
\label{eq:saddlepoints}
\begin{align}
&\nabla_{\bf k} S = \Delta {\bf x}_{\rm c} - \Delta {\bf x}_{\rm v} = 0 & \label{sp1} \\
&{dS \over dt'} = \varepsilon_{\rm g}({\bf k} - {\bf A}(t) + {\bf A}(t'))=0 & \label{sp2} \\
&{dS \over dt} = \varepsilon_{\rm g}({\bf k}) = n\omega\text{.} \label{sp3} &
\end{align}
\end{subequations}
where $S({\bf k},t',t) = \int_{t'}^{t} \varepsilon_{\rm g}[{\bf k} + {\bf A}(t') - {\bf A}(t)] d\tau$ is the action and ${\bf A}(t) = -\mathbf{A}_0 \sin{\omega t}$ is the vector potential of the driver. As discussed in Ref. \cite{vampa2014theoretical,vampa2015semiclassical}, Eqs. \ref{eq:saddlepoints} justify the semiclassical model of high-harmonic generation: Eq. (\ref{eq:saddlepoints}b) dictates that the newly created electron-hole pair accelerates in k-space according to the equation of motion ${\bf k} = {\bf A}(t') + {\bf A}(t)$; Eq. (\ref{eq:saddlepoints}a) requires the electron to re-encounter the hole; Eq. (\ref{eq:saddlepoints}c) determines that the high-harmonic photon energy equals the bandgap at the momentum of collision. 

In semiconductors and dielectrics, Eq. (\ref{eq:saddlepoints}b) has only a complex solution, a hallmark of quantum effects. So far, however, the saddle point equations in solids have been solved by zeroing the minimum bandgap, allowing real-valued solutions, effectively discarding the quantum contribution \cite{vampa2015semiclassical}. Here, instead, we solve Eqs. (\ref{eq:saddlepoints}) with a non-zero minimum bandgap, and verify the role that quantum corrections play in our experiment. The method follows closely that of Ref. \cite{ivanov2014multielectron} for gases. The complication arises from the non-parabolic band dispersion of crystals, parametrized here as $\varepsilon_g(\mathbf{k}) = E_g + \sum_{j=1}^N \Delta_j [1-\cos{(j\mathbf{k}a)]}$, where $k_b a = \pi$ ($k_b$ is the Bloch wavevector of the lattice). Only the band dispersion along the laser polarization ($\Gamma-X$) is considered. First, Eq. (\ref{eq:saddlepoints}b) is written as follows:
\begin{align}
    \Re &: \sum_j \Delta_j [1-\cos{(jx)}\cosh{(jy)}] = -E_g \label{eq:re}\\ 
    \Im &: \sum_j \Delta_j \sin{(jx)}\sinh{(jy)} = 0, \label{eq:im}
\end{align}
where $x/a = p'_{st} - A_0 \sin{\phi'_i}\cosh{\phi''_i}$, $y/a = p''_{st} - A_0\cos{\phi'_i}\sinh{\phi''_i}$, and $p_{st} = p'_{st} + ip_{st}''$ and $\phi_i' = \omega t_i = \phi_i'+i\phi_i''$ are the stationary canonical momentum and ionization phases respectively (to be determined). Eq. (\ref{eq:im}) is solved for either $x,y=0$. However, if $y = 0$, Eq. (\ref{eq:re}) has no solution because $\sum_j \Delta_j\cos{(jx)} \leq \sum_j \Delta_j < \sum_j \Delta_j + E_g > 0$ when all $\Delta_j \geq 0$ as in MgO (see caption of Supplementary Figure \ref{fig:trajectories}), in contradiction of Eq. (\ref{eq:re}). Therefore, $x=0$. Setting $\xi = \cosh{y}$ and expanding $\cosh{(jy)}$ in powers of $\xi$, Eq. (\ref{eq:re}) becomes a high-order polynomial whose real root $\alpha$ can be found numerically or, as is the case here, analytically for $N = 3$. Then, $p'_{st}, p''_{st}$ are expressed as:
\begin{subequations}
\label{eq:ps}
\begin{align}
    &p'_{st}(\phi_i',\phi_i'') - A_0\cos{\phi'_i}\sinh{\phi''_i} = \frac{\arcosh{\alpha}}{a}\\
    &p''_{st}(\phi_i',\phi_i'') - A_0\sin{\phi'_i}\cosh{\phi''_i} = 0
\end{align}
\end{subequations}
The same procedure is adopted for Eq. (\ref{eq:saddlepoints}c) to express:
\begin{subequations}
\label{eq:ps2}
\begin{align}
    &p'_{st}(\phi_r',\phi_r'') - A_0\sin{\phi'_r}\cosh{\phi''_r} = \frac{\arccos{\beta}}{a}\\
    &p''_{st}(\phi_r',\phi_r'') - A_0\cos{\phi'_r}\sinh{\phi''_r} = 0
\end{align}
\end{subequations}
where $\beta$ is the real root of the polynomial equation. In this case, choosing $x=0$ solves for below-gap harmonics, whereas $y=0$ solves for $n\omega \geq E_g$, which is what is needed in this work. Equations (\ref{eq:ps}) and (\ref{eq:ps2}) are identical to those found in the gas case \cite{ivanov2014multielectron} if one defines $\gamma = \arcosh\alpha/a, \gamma_N = \arccos\beta/a$. From Eqs. (\ref{eq:ps}) and (\ref{eq:ps2}) one derives:
\begin{subequations}
\label{phis}
\begin{align}
    \phi_i'(\phi_r',\phi_r'') = \arcsin{\sqrt{\frac{P-D}{2}}}\\
    \phi_i''(\phi_r',\phi_r'') = \arcosh{\sqrt{\frac{P+D}{2}}}
\end{align}
\end{subequations}
with $P = (p_{st}'/A_0)^2 + \tilde\gamma^2+1$, $D = \sqrt{P^2-4(p_{st}'/A_0)^2}$, and $\tilde\gamma = \gamma + p_{st}''/A_0$. Finally, Eqs. (\ref{eq:ps2}) and (\ref{phis}) are substituted in Eq. (\ref{eq:saddlepoints}a) and the integral is evaluated numerically for every harmonic photon energy $\Omega$ on a grid of $(\phi_r',\phi_r'')$. The minimum of $\lvert\nabla_k S\rvert^2$ determines the real and imaginary components of the recollision phase $\phi_r(\Omega)$, from which $\phi_i(\Omega)$ and $p_{st}(\Omega)$ can be calculated. \\

\begin{figure}
    \centering
    \includegraphics{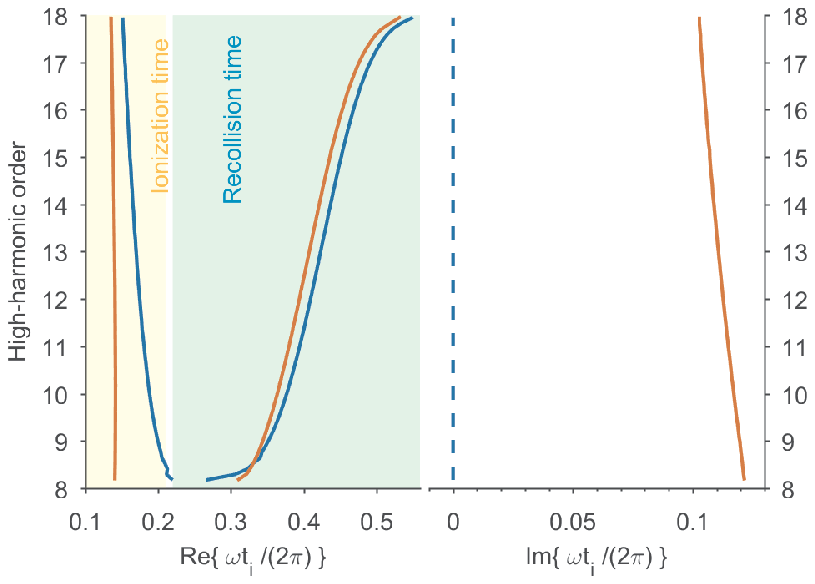}
    \caption{Comparison between classical (blue lines) and quantum (red lines) orbits. (left) Real part of the times of ionization and recombination; (right) imaginary part of the time of ionization. The classical model, neglecting quantum tunneling, predicts instantaneous ionization. The calculation is performed with: $F_0 = 0.63$  V/\AA, $~\omega = 1.3$ $\mu$m, $a = 4.21$ \AA, $E_g = 7.78$ eV, and $\Delta_1 = 4.521$ eV, $\Delta_2 = 1.021$ eV, $\Delta_3 = 0.1221$ eV. The band parameters are obtained from a multi-cosine fit of the band structures reported in Ref. \cite{you2017}, along the $\Gamma-X$ direction (the same probed in the experiment).} 
    \label{fig:trajectories}
\end{figure}

The predicted high-harmonic photon energy as a function of the complex "ionization time'' of the electron-hole pair is shown in Fig. \ref{fig:trajectories} (red lines), where it is also compared to the classical prediction (solving with $E_g = 0$, blue lines). Indeed, there are significant differences induced by the strong-field excitation step.

The quantum trajectories are used to calculate the phase between the fundamental and its second harmonic that maximizes the high-harmonic power, using the definition of $\sigma$ and the procedure reported in the previous section, replacing the real stationary points with the complex saddle points found here.

\section{Calculation of attochirp}
\begin{figure}
    \centering
    \includegraphics{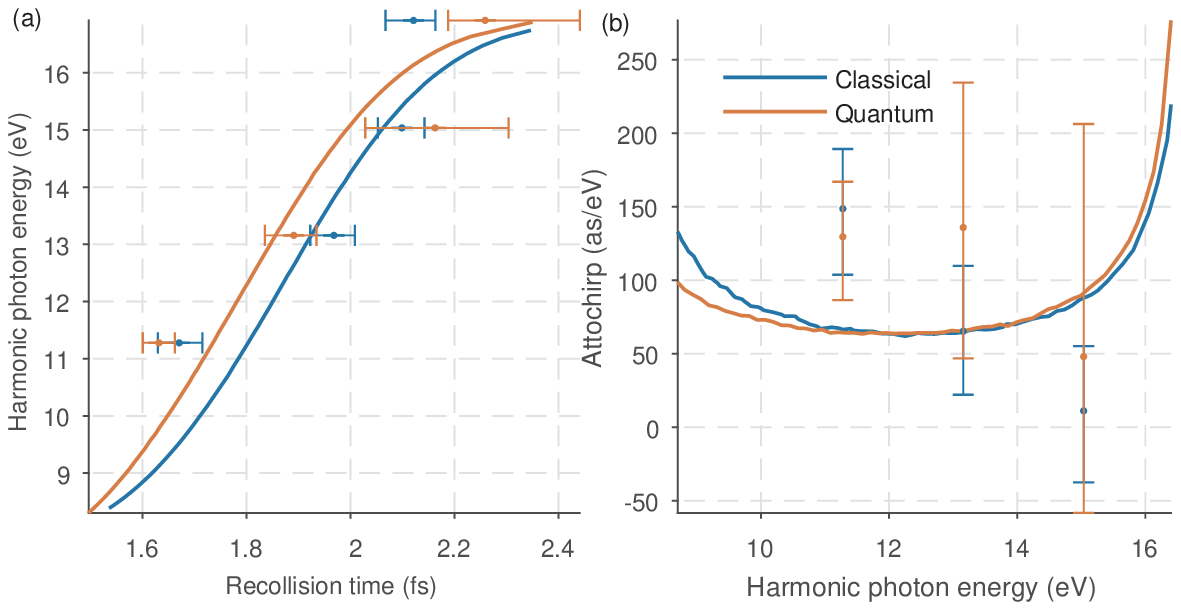}
    \caption{\textbf{(a)} Recollision time for all harmonics below or at the maximum bandgap, as mapped from the measured $\phi_{opt}$ with the classical (blue) and quantum (red) models. \textbf{(b)} Derivative of the recollision time with respect to harmonic photon energy (atto-chirp). The last point at $\sim 17$ eV is excluded because the next harmonic lies above the maximum bandgap and therefore the model doesn't account for it.}
    \label{fig:attochirp1}
\end{figure}

The emission time of each harmonic is extracted from the data presented in Fig. 3 of the main manuscript following the same procedure as in Ref. \cite{dudovich2006measuring}: for each even harmonic order the measured optimum phase $\phi_{opt}$ is interpolated to that predicted by the model, which links each $\phi_{opt}$ to a specific recollision time. The result is shown in Fig. \ref{fig:attochirp1}\textbf{a}. The attochirp is calculated as ${\Delta t_r}\over{\Delta\hbar\omega}$. It is shown in Fig. \ref{fig:attochirp1}\textbf{b}. Because the bandstructure of MgO that we employ does not extend beyond the 18$^{th}$ harmonic,  the semiclassical model does not allow harmonics beyond the 18$^{th}$. The quantum model is in principle capable of predicting harmonics beyond the classically-allowed region\cite{ivanov2014multielectron}, but those solutions have not been considered in this work. For both reasons, the emission time for the 20$^{th}$ harmonic is excluded in Fig. \ref{fig:attochirp1}(a). Because the attochirp about the 18$^{th}$ harmonic requires the emission time of the 20$^{th}$, it is excluded in the attochirp calculation.

\section{Bloch emission in two-color fields}
In this section we demonstrate that Bloch-type emission arising from independent acceleration of electrons and holes in their respective bands predicts no atto-chirp, contrary to the experimental observation.

For simplicity, we assume a 1-dimensional nearest-neighbor momentum-dependent band gap, $\varepsilon_g(k) = E_g + \Delta[1-\cos(ka)]$, where $a$ is the lattice constant. The difference in band velocities is $v_g(k) = \nabla_k\varepsilon_g = \Delta a \sin(ka)$. The laser field is $F(t) = F_1 \cos(\omega t) + F_2 \cos(2\omega t+\phi)$, where $F_1$ is the (strong) fundamental field strength and $F_2$ is the (weak) perturbation at the second harmonic of the fundamental field frequency. $\phi$ is the relative delay between the two colors. The crystal momentum of the electron-hole pair is time-dependent: $k(t) = -A_1(t) -A_2(t)= A_1 \sin(\omega t) +A_2 \sin(2\omega t + \phi)$, where $A_{1,2}(t)$ is the laser vector potential, $A_1 = F_1/\omega$, $A_2 = F_2/(2\omega)$. The time-dependent electron-hole velocity, then, is:
\begin{align}
    v_g(t) &= \Delta a \sin[A_1(t)+A_2(t)] =\\
    &= \Delta a \left[\sin{A_1(t)}\cos{A_2(t)}+\cos{A_1(t)}\sin{A_2(t)}\right] =\\
    &\simeq \Delta a \left[\sin{A_1(t)} + A_2(t)\cos{A_1(t)}\right].
\end{align}{}
where in the last step we assumed $A_2(t)<<1$ (perturbative regime). The first term in square brackets is the unperturbed spectrum (independent of the second-harmonic field), while the second term is the perturbed spectrum. The unperturbed spectrum rightfully comprises only the odd harmonics, and therefore will not be considered in the analysis of the even harmonics:
\begin{align}
    \sin{A_1(t)} &= \sin(A_1\sin\varphi_1) = 2\sum_{n=0}^\infty J_{2n+1}(aA_1)\sin[(2n+1)\varphi_1]
\end{align}{}
where we used the Jacobi-Anger expansion, and set $\varphi_1 = \omega t$. The perturbed spectrum comprises only even-order harmonics:
\begin{align}
    A_2(t)\cos{A_1(t)} &= A_2\sin\varphi_2\cos(A_1\sin\varphi_1) = \\
    &= A_2\sin\varphi_2 \left\{J_0(aA_1) + 2 \sum_{n=1}^\infty J_{2n}(aA_1)\cos(2n\varphi_1)\right\}
\end{align}
where we defined $\varphi_2 = 2\omega t + \phi$. The perturbation adds sidebands at $(2n\pm 2)\omega$, as we demonstrate below:
\begin{align}
    &= J_0(aA_1)A_2(t) + \sum_{n=1}^\infty J_{2n}(aA_1)A_2\left[ \sin(2n\varphi_1+\varphi_2)-\sin(2n\varphi_1-\varphi_2)\right]= \\
    &= J_0(aA_1)A_2(t) + \sum_{n=1}^\infty J_{2n}(aA_1)A_2\left[\sin((2n+2)\omega t+\phi)-\sin((2n-2)\omega t-\phi)\right]
\end{align}{}
Rearranging the sum:
\begin{align}
    v_g^{even}(t) &= \Delta a \left\{J_0(aA_1)A_2(t) + \sum_{m=1}^\infty A_2\left[J_{2m-2}(aA_1)\sin(2m\omega t+\phi)-J_{2m+2}(aA_1)\sin(2m\omega t-\phi)\right]\right\}
\end{align}{}

Finally, we calculate the spectrum of the even harmonics higher than the 2$^{nd}$ and their power as a function of $\phi$. For clarity we omit the factor $\Delta a A_2$ and the arguments of the Bessel functions ($aA_1$):
\begin{align}
    v_g(2N) &= FT\Big\{J_{2N-2}\sin(2N\omega t + \phi) - J_{2N+2}\sin(2N\omega t - \phi)\Big\} =\\
    &= FT\Big\{\sin(2N\omega t)\left[J_{2N-2}-J_{2N+2}\right]\cos(\phi) + 
    \cos(2N\omega t)\left[J_{2N-2}+J_{2N+2}\right]\sin(\phi)\Big\} =\\
    &= \Big\{\frac{1}{2i}\left[J_{2N-2}-J_{2N+2}\right]\cos(\phi) + \frac{1}{2}\left[J_{2N-2}+J_{2N+2}\right]\sin(\phi)\Big\}
    \delta(\Omega-2N).
\end{align}{}
Defining $A = J_{2N-2}-J_{2N+2}$, $B = J_{2N-2}+J_{2N+2}$, the even-harmonic power is:
\begin{align}
    I(2N) = |v_g(2N)|^2 &= \frac{1}{4}[A^2\cos^2\phi + B^2\sin^2\phi].
\end{align}{}
 Therefore, unless if $A=B$, the even harmonics modulate with $\phi$, but they all modulate with either the same phase or out-of-phase, depending whether $A > B$ or $A<B$. Bloch emission does not allow the continuosly varying modulation phase observed in the experiment.

\section{Dependence on 2$\omega$ power}
\begin{figure}
    \centering
    \includegraphics{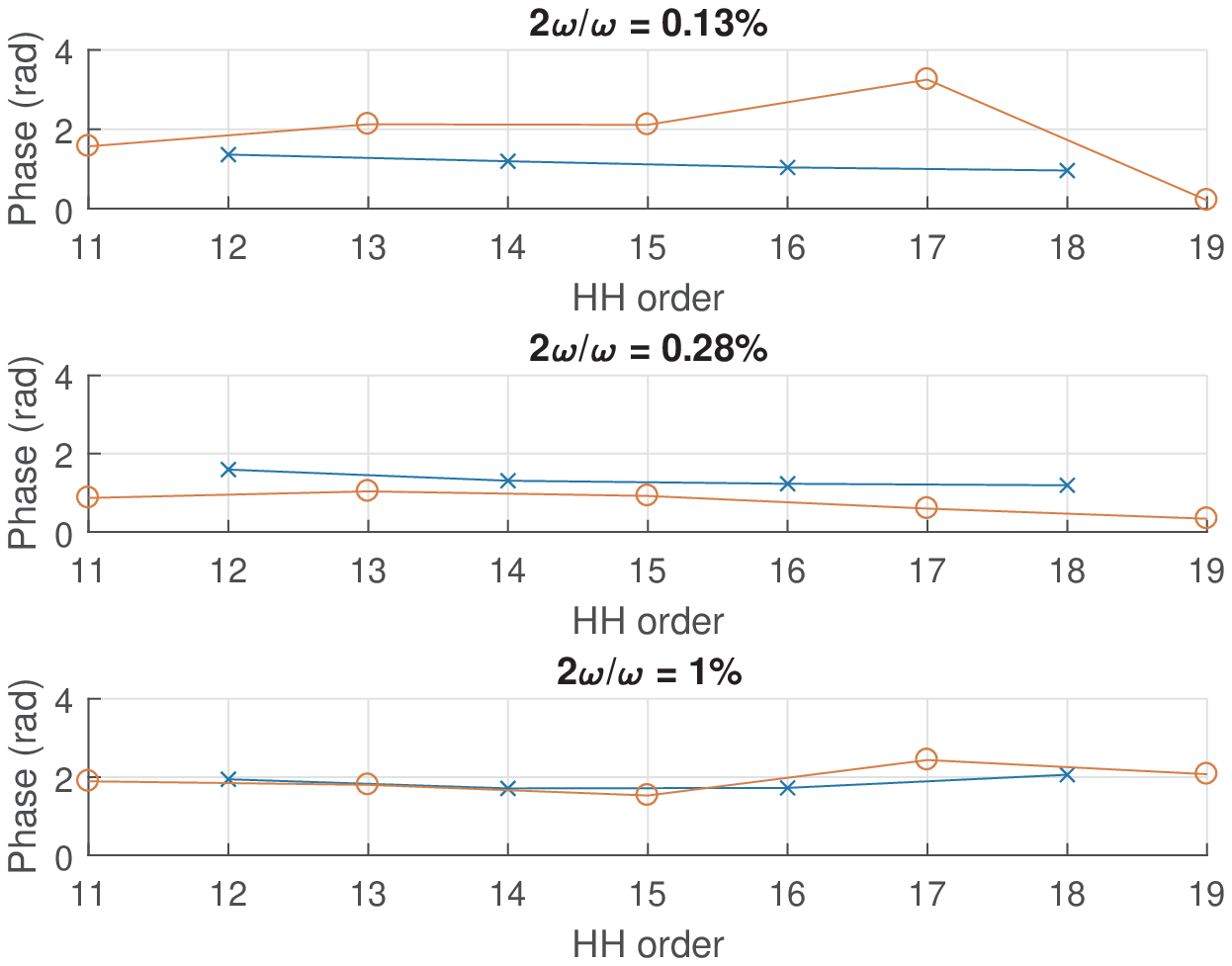}
    \caption{Modulation phase for even and odd harmonics versus second-harmonic power (relative to the fundamental power). Increasing $2\omega$ power yields progressively in-phase modulation of even and odd harmonics. The relative phase between adjacent even harmonics, though, changes little at the lower powers.}
    \label{fig:2wdep}
\end{figure}

Supplementary Figure \ref{fig:2wdep} reports the modulation phase of even and odd harmonics for three second-harmonic powers (relative to the fundamental power). Increasing the second-harmonic power reduces the relative modulation phase between even and odd harmonics, as well as a reduction of the modulation phase between the even harmonics (which we use to extract the "atto-chirp"). As a result, the extracted atto-chirps are to be regarded as lower bounds.
{}

\bibliography{biblio}